\documentclass[12pt,preprint]{aastex}
\usepackage[usenames]{color}

\shorttitle{The power-law distribution of flare kernels}
\shortauthors{Nishizuka et al.}

\begin{document}

\title{The Power-Law Distribution of Flare Kernels and Fractal Current Sheets in a Solar Flare}


\author{N. Nishizuka\altaffilmark{1}, A. Asai\altaffilmark{2,3}, H. Takasaki\altaffilmark{1,4}, H. Kurokawa\altaffilmark{1} and K. Shibata\altaffilmark{1}
}


\altaffiltext{1}{Kwasan and Hida observatories, Kyoto University, Yamashina, Kyoto 607-8471, Japan; nisizuka@kwasan.kyoto-u.ac.jp}
\altaffiltext{2}{Nobeyama Solar Radio Observatory, National Astronomical Observaotry of Japan, Minamisaku, Nagano 384-1305, Japan}
\altaffiltext{3}{The Graduate University for Advanced Studies (SOKENDAI), Miura, Kanagawa 240-0193, Japan}
\altaffiltext{4}{Accenture Japan, Ltd., Akasaka Inter City, Akasaka, Minato-ku, Tokyo 107-8672, Japan}

\begin{abstract}
We report a detailed examination of the fine structure inside flare ribbons and the temporal evolution of this fine structure during the X2.5 
solar flare that occurred on 2004 November 10. We examine elementary bursts of the $C_{IV}$ ($\sim$1550\AA) emission lines seen as local 
transient brightenings inside the flare ribbons in the ultraviolet (1600\AA) images taken with Transition Region and Coronal Explorer, and we 
call them $C_{IV}$ kernels. This flare was also observed in H$\alpha$ with the Sartorius 18 cm Refractor telescope at Kwasan observatory, 
Kyoto University, in hard X-rays (HXR) with Reuven Ramaty High Energy Solar Spectroscopic Imager. Many $C_{IV}$ kernels, whose sizes were 
comparable to or less than 2'', were found to brighten successively during the evolution of the flare ribbon. The majority of them were well 
correlated with the H$\alpha$ kernels in both space and time, while some of them were associated with the HXR emission. These kernels were 
thought to be caused by the precipitation of nonthermal particles at the foot points of the reconnecting flare loops. The time profiles of the 
$C_{IV}$ kernels showed intermittent bursts, whose peak intensity, duration and time interval were well described by power-law distribution 
functions. This result is interpreted as an avalanching system of ``self-organized criticality'' of a single event or by fractal current sheets 
in the impulsive reconnection region.
\end{abstract}


\section{ Introduction }

Hard X-ray (HXR) and microwave emissions show fine structures both temporally and spatially during a solar flare, which revealed that a 
highly-fragmented and intermittent particle acceleration occurs \citep[e.g.][]{ben92, asch02}. This fragmented structure of solar flares 
indicates that a flare is an ensemble of a vast amount of small scale energy release. Statistical studies of solar flares have also shown that 
various kinds of physical parameters of flares, like peak intensity, flare duration, waiting time of soft X-ray (SXR) emissions between discrete 
events, are well described with power-law distributions \citep[e.g.][]{den85, whe00, ver02}. Karlick\'{y} et al. (2000) examined twelve flares, 
and showed that microwave spikes seen in each flare show power-law features in size and time scales (i.e. scale of energy release). In addition, 
the occurrence of microflares and X-ray bright points is known to follow power-law distributions \citep{shim95, shimj99}.

Recent development in magnetic reconnection theory also indicates that magnetic reconnection proceeds intermittently, involving repeated 
formation of magnetic islands and their subsequence coalescence \citep{fin77, taj87}. This process is known as the ``impulsive bursty'' regime 
of magnetic reconnection \citep{pri85}. As Shibata $\&$ Tanuma (2001) showed, plasmoids of various scales are generated in the current sheet 
in a fractal manner. Such fractal nature of magnetic reconnection might generate power-law characteristics that are observed in solar flares, 
as mentioned above. Karlick\'{y} et al. (2000) and Kliem et al. (2000) discussed similar features seen in the HXR and microwave emissions, based 
on the theoretical view of dynamic magnetic reconnection. Although the temporal resolutions of HXR and microwave observations were high 
enough to reveal fragmented features in the temporal scale, the time variability of flare kernels has not been discussed with two-dimensional 
images with high spatial and temporal resolutions.
 
Historically, the two ribbon structure has been observed in H$\alpha$ and other wavelengths in solar flares. Flare kernels inside the ribbons are 
well correlated with HXR and microwave emissions temporally and spatially in H$\alpha$ \citep{kur88, kit90}. Also in ultraviolets (UVs), such as 
in 1550\,\AA\ taken with TRACE\footnote{Transition Region and Coronal Explorer \citep{han99}} the same properties were observed \citep{war01, 
ale06}, indicating that sudden plasma heating occurs in the upper chromosphere and the transition region by nonthermal particles or thermal 
conduction. Hence H$\alpha$ kernels and TRACE 1550\,\AA\ ($C_{IV}$ doublet emissions) kernels can also be good tracers of HXR sources.

In this paper, we examine the fine structures inside the flare ribbons seen in the UV images of the X2.5 flare that occurred on 2004 November 
10. We show the fragmented features of the bright emission sources, and that they follow a power-law distribution just in a single event. Finally 
we discuss the fractal features of energy release region (i.e. current sheet) and the avalanching system of the flare to explain such fragmented 
structures.

%
\section{Observations: Comparison among C$_{IV}$, H$\alpha$ and Hard X-ray Emissions}

The large flare (X2.5 in GOES class) occurred in the NOAA Active Region 10696 (N08$^{\circ}$, W50$^{\circ}$) at 02:00 UT, 2004 
November 10. This flare was a long duration event that showed a typical two ribbon structure preceded by a filament eruption. The 
erupted filament showed a kinking structure \citep{wil05}, and a lot of attention has been paid to it because this is a candidate for 
the source of the geo-effective coronal mass ejection \citep[CME;][]{har07}. We observed the flare with the Sartorius 18 cm 
Refractor Telescope at Kwasan Observatory, Kyoto University \citep{asa03}. The highest temporal and spatial resolutions of the 
Sartorius data are 1 $s$ and 1''.2, respectively. Figures 1(a)-1(c) shows the images of the flare in H$\alpha$ at 02:06, 02:08, and 
02:10 UT, which correspond to the peak times of the HXR emission (see also Figure 2). We can see some H$\alpha$ flare kernels 
inside the ribbon structure. 

The UV images were taken by 1600\,\AA\ passband of TRACE as shown in Figures 1(d)-1(f). They also show a two-ribbon structure. 
The TRACE 1600\,\AA\ data was obtained with the temporal and the spatial resolutions of 3 $s$ and 0.5''. During flares, the pair of 
$C_{IV}$ doublet ($\sim$1550\AA) in a broad response of the 1600\,\AA\ passband is strongly enhanced \citep{bre96, war00}. The 
$C_{IV}$ formation temperature is about 10$^5$ K. Therefore, the bright features in Figures 1(d)-1(f) observed in the impulsive phase 
are probably caused by the enhancement of the $C_{IV}$ ($\sim$1550\AA) emission line due to transition region heating. We call the 
bright features $C_{IV}$ kernels.

We overlaid HXR contour images (25-50 keV) on TRACE 1600\,\AA\ images to compare the spatial distribution of radiation sources 
in H$\alpha$ and HXR emissions (see Figures 1(d)-1(f)). The HXR images were taken with RHESSI\footnote{Reuven Ramaty High Energy 
Solar Spectroscopic Imager \citep{lin02}}. We synthesized the HXR image with the Clean algorithm, which is the same method as is 
commonly used for analysis of radio data, and grids 3-9, which give the spatial resolution (FWHM) of about 10''. The integration time 
is set to be 60 $s$, and the total photon count was 3.8-7.7$\times$10$^{5}$ counts for photons of 25-50 keV. These synthesizing 
tools are included in the Solar SoftWare. We found that the HXR sources are associated with both H$\alpha$ and $C_{IV}$ kernels. 
The location of the HXR sources moves in the southeast direction as the flare progresses, that is, from a mixed polarity region to a 
strong magnetic field region, indicating a change in the site of the strong energy release. Though the kernels are seen in the southeast 
of the H$\alpha$ and TRACE images from 2:05-2:08 i.e. before the HXR sources have arrived there, this is probably because the HXR 
emissions are not large enough to be observed with the dynamic range of RHESSI. Actually small flare kernels in the southeast of Figure 
1(d) at around 02:06 UT, which are the components of the ribbons, show small peaks of intensity less than 25 percent of the later 
impulsive burst at 02:10 UT, as well as in the H$\alpha$ time profile (see Figure 2(b)).

We summarize the results of the comparison of the multi-wavelength observations in H$\alpha$, $C_{IV}$ ($\sim$1550\AA) and HXR 
emissions as follows: (i) There are not only good spatial but also temporal correlations among flare kernels observed in H$\alpha$ and 
$C_{IV}$ emissions, some of which are associated with the HXR emission. This implies that the $C_{IV}$ and H$\alpha$ kernels are caused 
by nonthermal electrons interacting with the ambient thick target plasma as well as HXRs. (ii) The $C_{IV}$ flare ribbons are much thinner 
and sharper than the H$\alpha$ ribbons. This is because the width of flare ribbons is typically determined by cooling time via thermal 
conduction and radiative cooling, and because thermal conduction time scale in the corona/transition region for $C_{IV}$ is much shorter 
than that for H$\alpha$ in the chromosphere. The ratio of the peak intensity of the H$\alpha$ kernels to the background is not very large, 
as a result, the integrated H$\alpha$ emission over the whole active region is similar to the soft X-ray emission (see Figure 2(b)). On the 
other hand, the integrated $C_{IV}$ emission is still similar to the HXR emission, showing corresponding peaks in their time profiles.

%
%
\section{Analysis and Result}

As a result of the comparison of the multi-wavelength observations, we found that it is easier to identify peaks in the time profile of 
the $C_{IV}$ emission, rather than in H$\alpha$. Moreover, the seeing condition smeared the H$\alpha$ images in the impulsive phase, 
and therefore we focused on the temporal variations of the $C_{IV}$ flare kernels here. We measured the intensity, duration and time 
interval between each peak from the time profiles. We divided both flare ribbons into fine meshes. Each mesh box is a square with size 
of 5'', 2'' and 1'' for comparison. Although this is larger than the elemental H$\alpha$ kernels, which are considered to be about 1'' or 
even smaller \citep{kur86}, it is small enough for us to determine the essential structures inside the flare ribbons. Next we examined 
time variations of the total intensity for each box in the meshes. As the mesh size becomes smaller and smaller, peaks in the time profile 
become isolated. This means that the light curves with a large (5'') mesh possibly contain multiple flare kernels that are superposed over 
each other, while smaller size mesh can cover only single flare kernel. We found that a 2'' mesh is enough to isolate most of the superposed 
peaks, though some peaks cannot be separated even with a 1'' mesh. This implies that the size of the heating source is comparable to or 
smaller than 2''. Since a 1'' mesh is too small and too noisy to analyze, we adopted a 5'' and a 2'' size mesh for our further analysis.

We defined the maximum intensity of a light curve as the peak intensity ($I$), and determined the duration ($t_d$) as the full width at the 
three fourths maximum intensity of each peak because not all of the peak durations can be measured with the full width at the half maximum 
(FWHM). We identified 586 $C_{IV}$ kernels using a 5'' mesh in the impulsive phase, only with the requirement that the count rate of the 
detector exceeds 50 counts s$^{-1}$ to identify the peaks. Figure 3(a), 3(b) shows the frequency distributions of the peak intensity and the 
duration of each peak. We also recorded the peak time of the flare kernels across the whole active region. We determined the time interval 
of the peaks ($t_{int}$) as the time difference between the peak times and show its frequency distribution in Figure 3(c). The distribution of 
peak intensities, durations and time intervals reveal power-laws during the impulsive phase. From the slopes of the distribution, we obtain a 
power-law indices $\alpha$ $\sim$1.5 for the peak intensity, $\alpha$ $\sim$2.3 for the peak duration, and $\alpha$ $\sim$1.8 for the time 
interval between each peak. The lower limit of time duration of about 10 $s$ comes from the temporal resolution of TRACE such as 2-3 $s$ 
in a flare mode. When we change the mesh size from 5'' to 2'', each peak became isolated and sharpened so that the number of the peaks 
with short duration increased.

%
%
\section{Summary and Discussion}

We found that the distributions of the peak intensity, duration and time interval well followed power-law distributions with the power-law 
indices of $\alpha$ $\sim$1.5, 2.3 and 1.8, respectively. These power-law indices remain unchanged, even if we change the size of the mesh 
box from 5'' to 2'', and even if we change the threshold of the peak identification. In this individual event, we showed for the first time, the 
power-law behavior of flare kernels typically seen in studies of large numbers of flares, suggestive of a link between the observations and 
theoretical modeling of the fractal nature of magnetic reconnection in current sheets. If magnetic reconnection occurs in a fractal manner 
in the current sheet, one would expect energy release and particle heating/acceleration on a range of different sizes and time scales, such 
that power-law distributions could be expected in the size, duration, etc. of tracers of the energy release process. Since flare kernels have 
been shown to be good proxies for the HXR energy release and, furthermore, TRACE $C_{IV}$ kernels can also be good tracers of HXR sources, 
one would expect to see such behavior in their properties. In fact, the peak intensity and peak duration could be indicators of the released 
energy. The peak time also corresponds to the timing when heating of the foot point plasma occurs, that is the arrival time of released energy 
at the foot point. 

The duration of the transition region heating $t_d$ and the time interval $t_{int}$ are roughly characterized by Alfv\'{e}n time $t_A$ of the 
reconnection region, 
\begin{eqnarray}
t_A = L/v_A \propto L/B,
\end{eqnarray}
where $L$ is a characteristic length of the energy release region (e.g. macroscopic length of a current sheet or plasmoid), $v_A$ is an 
Alfv\'{e}n velocity ($\propto$ $B$), and $B$ is a typical magnetic field strength in the corona. On the other hand, intensity of flare kernels 
$I$ is estimated as,
\begin{eqnarray}
I \propto \frac{B^2}{t_A}L^3 \propto \frac{B^2}{L/B}L^3 = B^3L^2. 
\end{eqnarray}
So, if magnetic reconnection occurs in a fractal manner in the current sheet through the repeated formations of magnetic islands and their 
subsequence coalescence, current sheets become thinner and thinner and as a result, the self-similar structure of current sheet can be 
formed from macroscopic to microscopic scales. At that time, the size of energy release region $L$ can be expected to exhibit power-law 
behavior, so that power-law distributions can be expected in the energy, duration, etc. of tracers of the energy release process, such as $I$ 
and $t_A$.

These fractal structures mean that there are no characteristic scales of length, energy and time in the energy release process. Our results 
also support the view of the impulsive bursty reconnection \citep{pri85} and the fractal features of the current sheet \citep{shi01}. A power-law 
distribution for magnetic energy of plasmoid is also reported in the magnetosphere by Hoshino et al. (1994). On the basis of the unified view 
suggested by Shibata (1999) as the plasmoid-induced-reconnection model, plasmoid ejection plays a crucial role for energy storage and release, 
driving the inflow and the reconnection rate enhancement. On a large scale, the flare itself should exhibit these properties. Our results are 
quite similar to the power-law behaviors typically seen in studies of large numbers of flares \citep[e.g.][]{den85}, which are often interpreted 
as evidence of an avalanching system of self-organized criticality (SOC). This suggests that the elemental energy release in this individual 
event may be similar to that in a typical X-ray flare and hence interpreted as an avalanching system of SOC in a single event or by fractal 
current sheet in the impulsive reconnection region as discussed above. 

\acknowledgments

We first acknowledge an anonymous referee for his/her useful comments and suggestions. We wish  to acknowledge all the members of 
Kwasan Observatory for their support during our observation, especially M. Kamobe and  A. Edamura. We also thank A. Hillier for his careful 
reading and correction of this Letter. We would like to thank TRACE and RHESSI data center for their extensive use. This work was 
supported in part by the Grant-in-Aid for Creative Scientific Research ``The Basic Study of Space Weather Prediction'' (Head Investigator: 
K. Shibata) from the Ministry of Education, Culture, Sports, Science, and Technology of Japan, and in part by the Grand-in-Aid for the 
Global COE program ``The Next Generation of Physics, Spun from Universality and Emergence'' from the Ministry of Education, Culture, 
Sports, Science, and Technology (MEXT) of Japan.\\

\begin{figure}
\epsscale{.80}
\plotone{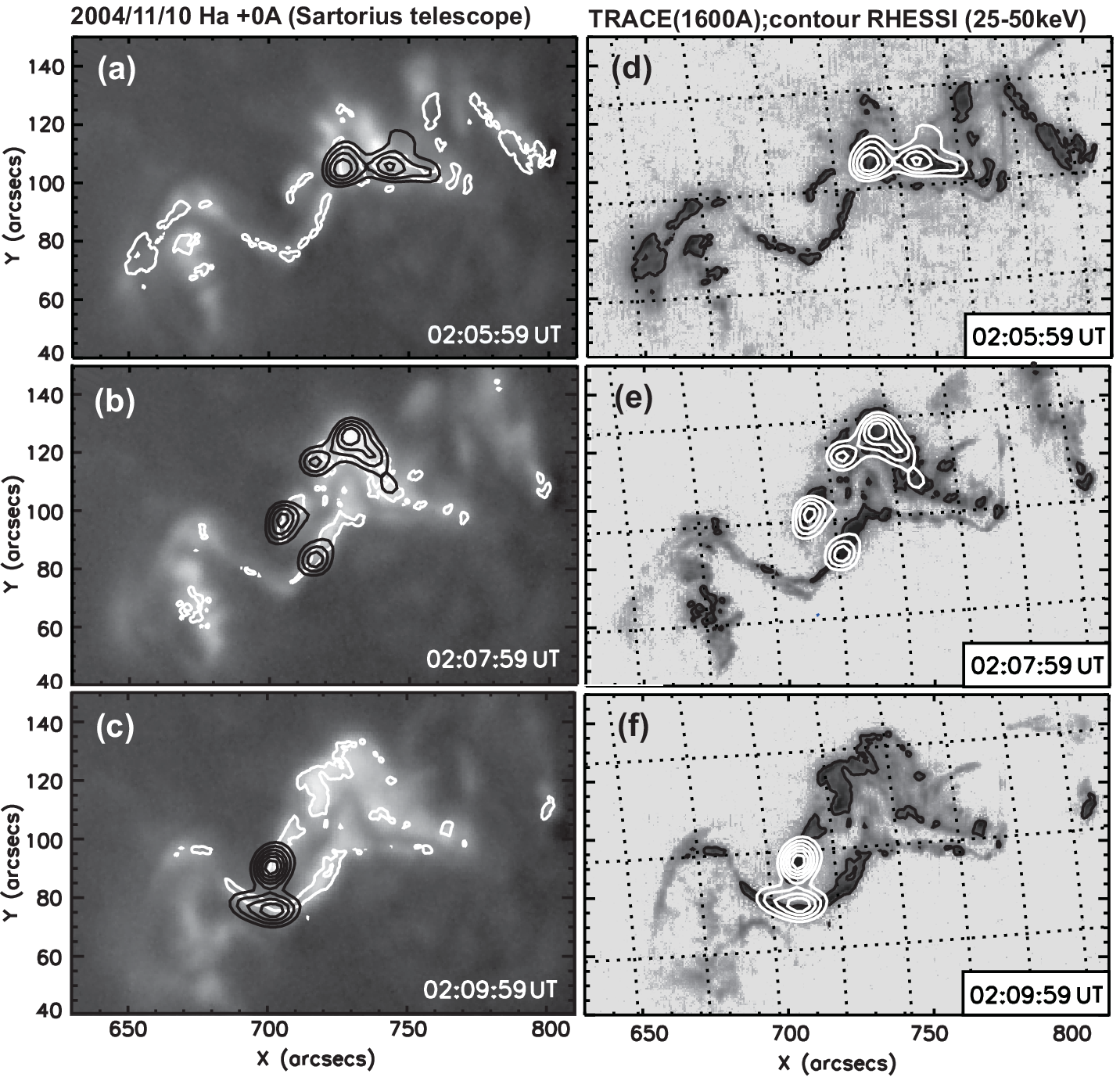}
\caption{Temporal evolution of the two-ribbon flare that occurred on 2004 November 10. (a)-(c) H$\alpha$ images taken with Sartorius at 2:06, 
2:08 and 2::0 UT, corresponding to peak times of the HXR emission (see Figure 2(a)). (d)-(f) Negative images of the TRACE 1600\,\AA\ data at 
the same times as the H$\alpha$ images. The HXR contour images taken with RHESSI (25-50 keV) are overlaid on them. We also overlaid a 
contour plot of $C_{IV}$ over H$\alpha$ images. \label{fig1}}
\end{figure}

\begin{figure}
\epsscale{.60}
\plotone{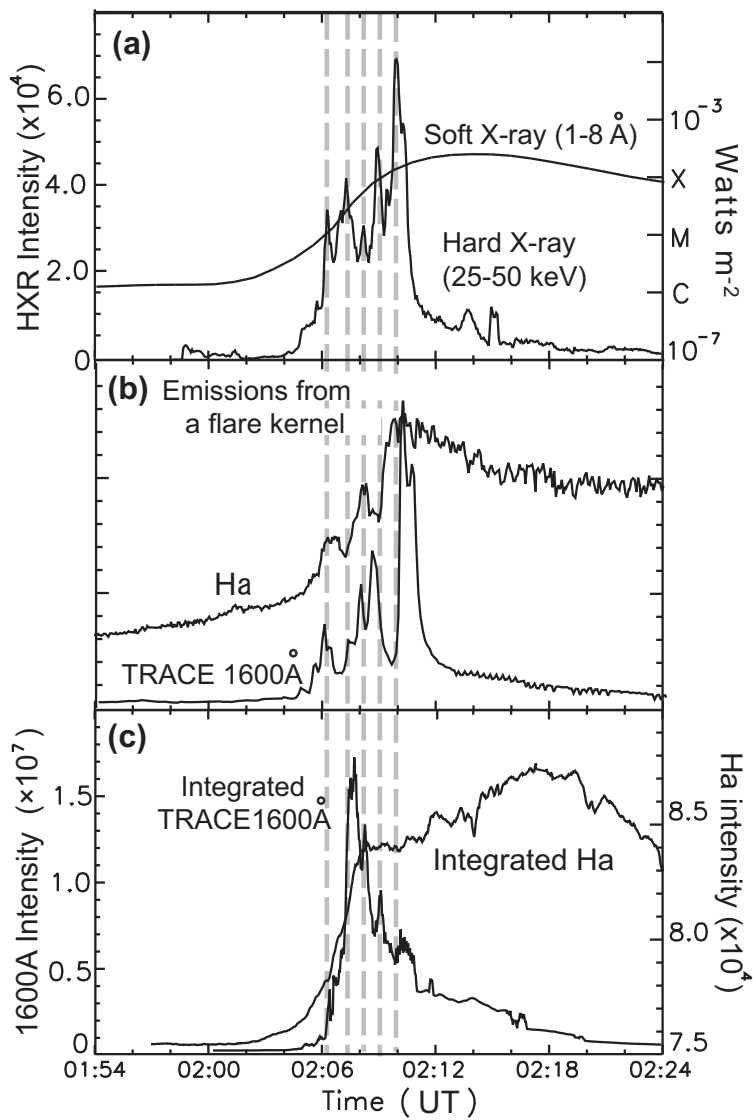}
\caption{Time profiles of multi-wavelength emissions in the 2004 November 10 flare. (a) Total intensities of the HXR emission obtained with 
RHESSI (25-50 keV) and the soft X-ray emission with GOES. (b) The time profiles of flare kernels observed both in $C_{IV}$ and H$\alpha$ show 
spiky structures, synchronizing with HXR intermittent bursts. (c) Total intensities of $C_{IV}$ ($\sim$1550\AA) and H$\alpha$ ribbons, consisting 
of a gradual rise and spikes. \label{fig2}}
\end{figure}

\begin{figure}
\epsscale{.70}
\plotone{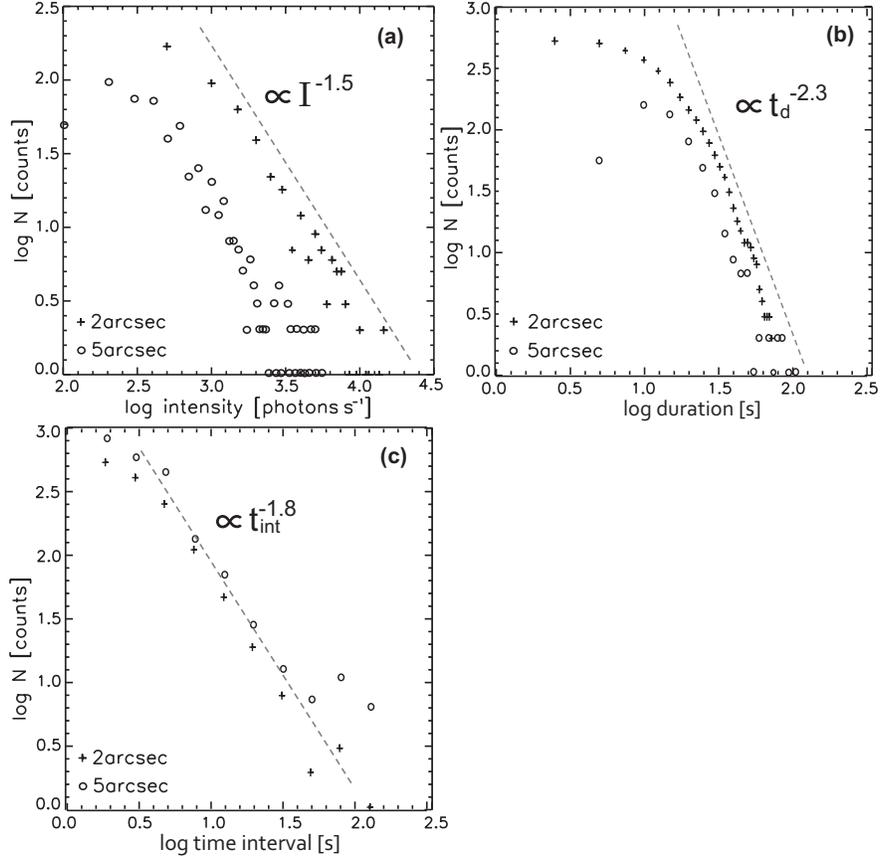}
\caption{Frequency distributions as function of (a) the peak intensity, (b) duration and (c) time interval of each burst in log-log space. Circle symbol 
shows the result in the case of mesh boxes with the size of 5'', and plus symbol shows that with 2''. All distributions can be approximated with power 
law functions through the impulsive phase. \label{fig3}}
\end{figure}

\clearpage




\begin{thebibliography}{}
%
\bibitem[Alexander \& Coyner (2006)]{ale06} Alexander, D. \& Coyner, A. J. 2006, \apj, 640, 505
%
\bibitem[Asai et al. (2003)]{asa03} Asai, A., Ishii, T. T., Kurokawa, H., Yokoyama, T., \& Shimojo, M., 2003, \apj, 586, 624
%
\bibitem[Aschwanden (2002)]{asch02} Aschwanden, M. J. 2002, Particle Acceleration and Kinematics in Solar Flares (Dordrecht: Kluwer)
%
\bibitem[Benz \& Aschwanden (1992)]{ben92} Benz, A. O. \& Aschwanden, M. J. 1992, in Lecture Notes in Physics, Vol. 399: Eruptive Solar Flares, 
eds. Z. Svestka, B. V. Jackson, \& M. E. Machado (New York: Springer), 106
%
\bibitem[Brekke et al. (1996)]{bre96} Brekke, P., Rottman, G. J., Fontenla, J. \& Judge, P. G., 1996, \apj, 468, 418
%
\bibitem[Dennis (1985)]{den85} Dennis, B. R. 1985, \solphys, 100, 465
%
\bibitem[Finn \& Kaw (1977)]{fin77} Finn, J. M. \& Kaw, P. K. 1977, Phys. Fluids, 20, 72
%
\bibitem[Handy et al. (1999)]{han99} Handy, B. N., et al. 1999, \solphys, 187, 229 
%
\bibitem[Harra et al. (2007)]{har07} Harra L. K., et al. 2007, \solphys. 244, 95
%
\bibitem[Hoshino et al. (1994)]{hos94} Hoshino, M., Nishida, A., Yamamoto, T., \& Kokubun, S., 1994, Geophys. Res. Let., 21, 25, 2935
%
\bibitem[Karlick\'{y} et al. (2000)]{kar00} Karlick\'{y}, M., Ji\v{r}i\v{c}ka, K. \& Sobotka, M. 2000, \solphys, 195, 165
%
\bibitem[Kitahara \& Kurokawa (1990)]{kit90} Kitahara, T. \& Kurokawa, H. 1990, \solphys, 125, 321
%
\bibitem[Kliem et al. (2000)]{kli00} Kliem, B., Karlick\'{y}, M. \& Benz, A. O. 2000, A\&A, 360, 715
%
\bibitem[Kurokawa (1986)]{kur86} Kurokawa, H. 1986, in Proc. of NSO/SMM Flare Symp., Low Atmosphere of Solar Flares, ed. D. Neidig (Sunspot:NSO), 51
%
\bibitem[Kurokawa et al. (1988)]{kur88} Kurokawa, H., Takahara, T. \& Ohki, K. 1988, Publ. Astron. Soc. Japan, 40, 357
%
\bibitem[Lin et al. (2002)]{lin02} Lin, R. P., et al. 2002,	\solphys, 210, 3
%
\bibitem[Priest (1985)]{pri85} Priest, E. R., 1985, Rep. Prog. Phys. 48, 955 
%
\bibitem[Shibata (1999)]{shi99} Shibata, K. 1999, Astrophys. Sp. Sci., 264, 129 
%
\bibitem[Shibata \& Tanuma (2001)]{shi01} Shibata, K. \& Tanuma, S. 2001, Earth, Planets and Space, 53, 473
%
\bibitem[Shimizu et al. (1995)]{shim95} Shimizu, T. 1996, Publ. Astron. Soc. Japan, 47, 251
%
\bibitem[Shimizu et al. (2008)]{shim08} Shimizu, M., et al. 2008, \apj, 683, L203 
%
\bibitem[Shimojo \& Shibata (1999)]{shimj99} Shimojo, M. \& Shibata, K. 1999, \apj, 516, 934
%
\bibitem[Tajima et al. (1987)]{taj87} Tajima, T., Sakai, J., Nakajima, H., Kosugi, T., Brunel, F., Kundu, M. R., 1987, \apj, 321, 1031
%
\bibitem[Veronig et al. (2002)]{ver02} Veronig, A., Temmer, M., Hanslmeier, A., Otruba, W. \& Messerotti, M., 2002, A\&A, 382, 1070
%
\bibitem[Warren \& Winebarger (2000)]{war00} Warren, H. P. \& Winebarger, A. R., 2000, \apj, 535, L63
%
\bibitem[Warren \& Warshall (2001)]{war01} Warren, H. P. \& Warshall, A. D. 2001, \apj, 560, L87
%
\bibitem[Wheatland (2000)]{whe00} Wheatland, M. S. 2000, \apj, 536, L109
%
\bibitem[Williams et al. (2005)]{wil05} Williams, D. R., T\"{o}r\"{o}k, T., D\'{e}moulin, P., van Driel-Gesztelyi, L., Kliem, B., 2005, \apj, 628, L163
%
%
%
\end{thebibliography}
\end{document}